# From Fields to Trees

**Firas Hamze** **Nando de Freitas**

Computer Science Department
University of British Columbia
{fhamze,nando}@cs.ubc.ca

## Abstract

We present new MCMC algorithms for computing the posterior distributions and expectations of the unknown variables in undirected graphical models with regular structure. For demonstration purposes, we focus on Markov Random Fields (MRFs). By partitioning the MRFs into non-overlapping trees, it is possible to compute the posterior distribution of a particular tree exactly by conditioning on the remaining tree. These exact solutions allow us to construct efficient blocked and Rao-Blackwellised MCMC algorithms. We show empirically that tree sampling is considerably more efficient than other partitioned sampling schemes and the naive Gibbs sampler, even in cases where loopy belief propagation fails to converge. We prove that tree sampling exhibits lower variance than the naive Gibbs sampler and other naive partitioning schemes using the theoretical measure of maximal correlation. We also construct new information theory tools for comparing different MCMC schemes and show that, under these, tree sampling is more efficient.

## 1 INTRODUCTION

Rao-Blackwellised sampling is a powerful inference tool for probabilistic graphical models (Doucet, de Freitas, Murphy and Russell 2000, Paskin 2003, Bidyuk and Dechter 2003). In this paper, we propose a new Rao-Blackwellised MCMC algorithm for MRFs, which is easily expandable to other models, such as conditional random fields (Kumar and Hebert 2003, McCallum, Rohanimanesh and Sutton 2003). MRFs play an important role in spatial statistics and computer vision (Besag 1986, Besag 1974, Li 2001). Existing MCMC algorithms for MRFs tend to be slow and fail to exploit the structural properties of the MRF graphical model (Geman and Geman 1984, Swendsen and Wang 1987). In contrast, variational approximation schemes (Yedidia, Freeman and Weiss 2000, Wainwright, Jaakkola and Willsky 2003) do exploit structural properties, but may often fail to converge.

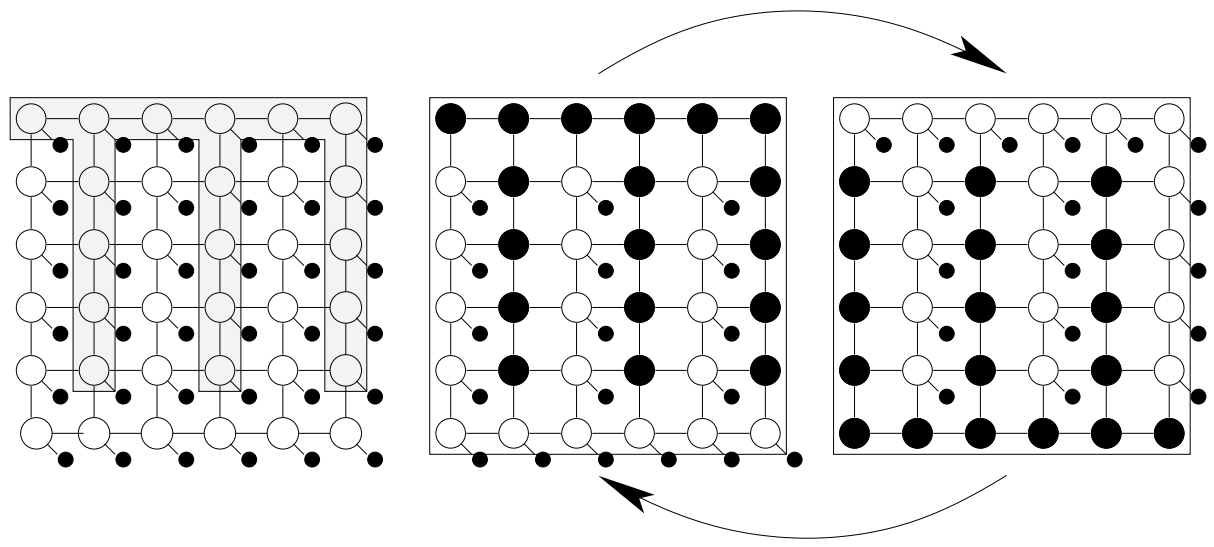

Figure 1: At left, illustration of a partitioned MRF; nodes in the shaded and white regions are $\Delta_1, \Delta_2$ respectively, with the small black circles representing observations. At right, depiction of the two-stage sampler; *sampled* values are *large* black circles. Conditioned on $\Delta_1$, the variables in $\Delta_2$ form a tree. Using this two-stage scheme, Rao-Blackwellised estimators are guaranteed to outperform naive ones.

The algorithm proposed in this paper exploits the property that MRFs can be split into two disjoint trees. By carrying out exact inference on each tree, it is possible to sample half of the MRF nodes in a single MCMC step. Our theorems will show that this tree sampling method outperforms simpler MCMC schemes. In particular, it exhibits lower correlation between samples and a faster geometric convergence rate. These theoretical results will be mirrored by our numerical examples.

## 2 TREE SAMPLING FOR MRFS

### 2.1 MODEL SPECIFICATION

We specify the MRF distribution on a graph $\mathcal{G}(\mathcal{V}, \mathcal{E})$, with edges $\mathcal{E}$ and $N$ nodes $\mathcal{V}$ as shown in Figure 1 left. The clear nodes correspond to the unknown discrete states $\mathbf{x} \in \{1, \ldots, n_x\}$, while the attached black nodes represent discrete observations $\mathbf{y} \in \{1, \ldots, n_x\}$ (they could also be Gaussian). According to this graph, the

MRF distribution factorizes into a product of local Markov positive potentials:

$$P(\mathbf{x}_{1:n}, \mathbf{y}_{1:n}) = \frac{1}{Z} \prod_{i \in \mathcal{V}} \phi(\mathbf{x}_i, \mathbf{y}_i) \prod_{(i,j) \in \mathcal{E}} \psi(\mathbf{x}_i, \mathbf{x}_j)$$

where $Z$ is the partition function, $\phi(\cdot)$ denotes the observation potentials and $\psi(\cdot)$ denotes the pair-wise interaction potentials. Our goal is to estimate the marginal posterior distributions (beliefs) $p(\mathbf{x}_i|\mathbf{y}_{1:N})$ and expectations of functions over these distributions.

As shown in Figure 1, *an MRF can be partitioned into two disjoint trees.* The loops in the MRF model cause it to be analytically intractable. However, belief propagation on each of the two trees is a tractable problem. This idea leads naturally to an algorithm that combines analytical and sampling steps. In particular if we have a sample of one of the trees, we can use belief propagation to compute the exact distribution of the other tree by conditioning on this sample. The algorithm therefore alternates between computing the trees and sampling them, as shown in Figure 1. Drawing samples in blocks (trees in this case) is well known to have benefits over algorithms that sample one node at-a-time. In Section 3 we will prove the domination of estimators using this tree sampling algorithm in comparison to other sampling schemes. Before this, we present the algorithm in more detail.

### 2.2 TREE SAMPLING ALGORITHM

Consider the 5x5 MRF graph shown in Figure 1 at left. We have partitioned the nodes into two disjoint sets. Denote the set of indices of the shaded nodes as $\Delta_1$ and those of the unshaded nodes as $\Delta_2$, where of course, $\Delta_1 \cup \Delta_2 = I$, the set of all node indices. Let the variables indexed by these nodes be $X_{\Delta_1} \triangleq \{X_j | j \in \Delta_1\}$ and $X_{\Delta_2} \triangleq \{X_j | j \in \Delta_2\}$. If we can sample from the conditional distributions:

$$\begin{array}{c} p(\mathbf{x}_{\Delta_1}|\mathbf{x}_{\Delta_2}, \mathbf{y}) \\ p(\mathbf{x}_{\Delta_2}|\mathbf{x}_{\Delta_1}, \mathbf{y}) \end{array} \quad (1)$$

then we have a two-stage Gibbs sampler called *data augmentation*, which has powerful structural properties that the general Gibbs sampler lacks (Liu 2001, Robert and Casella 1999). In particular, the two Markov chains in data augmentation exhibit the *interleaving property*: $\mathbf{x}_{\Delta_2}^{(t)}$ is independent of $\mathbf{x}_{\Delta_2}^{(t-1)}$ given $\mathbf{x}_{\Delta_1}^{(t)}$; and $(\mathbf{x}_{\Delta_2}^{(t)}, \mathbf{x}_{\Delta_2}^{(t-1)})$ and $(\mathbf{x}_{\Delta_2}^{(t)}, \mathbf{x}_{\Delta_2}^{(t)})$ are identically distributed under stationarity.

Conditioned on set $X_{\Delta_2}$, the variables $X_{\Delta_1}$ form an acyclic graph whose marginals are readily computed using *belief propagation* (Pearl 1987). This enables us to sample efficiently from the joint conditionals in (1) using the *Forward Filtering/Backward Sampling* algorithm (FF/BS) (Carter and Kohn 1994, Wilkinson and Yeung 2001). The sampling cycle is graphically shown in Figure 1, which makes it explicit that sampled values act as additional "evidence" to the complementary graph. The algorithm is shown in pseudocode in Figure 2.

**Tree sampling**

Initialize

- Set all sums $S_i = 0$
- Partition the set of all nodes $I$ into disjoint, tree-connected sets $\{\Delta_1, \Delta_2\}$
- for $i \in I$, randomly initialize $X_i^{(0)}$

for $t = 1 \ldots T$

- for $i \in \Delta_1$
  - Apply belief propagation to compute the *smoothing* densities $p(\mathbf{x}_i|\mathbf{x}_{\Delta_2}^{(t-1)}, \mathbf{y})$, treating the samples $\Delta_2^{(t-1)}$ as evidence.
  - Compute the expectations for the Rao-Blackwellised estimator $\mathbb{E}[h(X_i)|\mathbf{x}_{\Delta_2}^{(t-1)}, \mathbf{y}] = \sum_{X_i} h(X_i) p(\mathbf{x}_i|\mathbf{x}_{\Delta_2}^{(t-1)}, \mathbf{y})$
  - Set $S_i \leftarrow S_i + \mathbb{E}[h(X_i)|\mathbf{x}_{\Delta_2}^{(t-1)}, \mathbf{y}]$
  - Sample $X_{\Delta_1}^{(t)} \sim p(\mathbf{x}_{\Delta_1}|\mathbf{x}_{\Delta_2}^{(t-1)}, \mathbf{y})$ using *Forward filtering / Backward sampling.*
- for $i \in \Delta_2$
  - Apply belief propagation to compute the *smoothing* densities $p(\mathbf{x}_i|\mathbf{x}_{\Delta_1}^{(t)}, \mathbf{y})$, treating the samples $\Delta_1^{(t)}$ as evidence.
  - Compute the expectations for the Rao-Blackwellised estimator $\mathbb{E}[h(X_i)|\mathbf{x}_{\Delta_1}^{(t)}, \mathbf{y}] = \sum_{X_i} h(X_i) p(\mathbf{x}_i|\mathbf{x}_{\Delta_1}^{(t)}, \mathbf{y})$
  - Set $S_i \leftarrow S_i + \mathbb{E}[h(X_i)|\mathbf{x}_{\Delta_1}^{(t)}, \mathbf{y}]$
  - Sample $X_{\Delta_2}^{(t)} \sim p(\mathbf{x}_{\Delta_2}|\mathbf{x}_{\Delta_1}^{(t)}, \mathbf{y})$ using *Forward filtering / Backward sampling.*

Output Rao-Blackwellised estimates

- $\delta_{rb}(h(X_i)) = \frac{1}{T} S_i$

Figure 2: Tree sampling.

In the pseudocode, we are adopting *Rao-Blackwellised* estimators (Casella and Robert 1996, Gelfand and

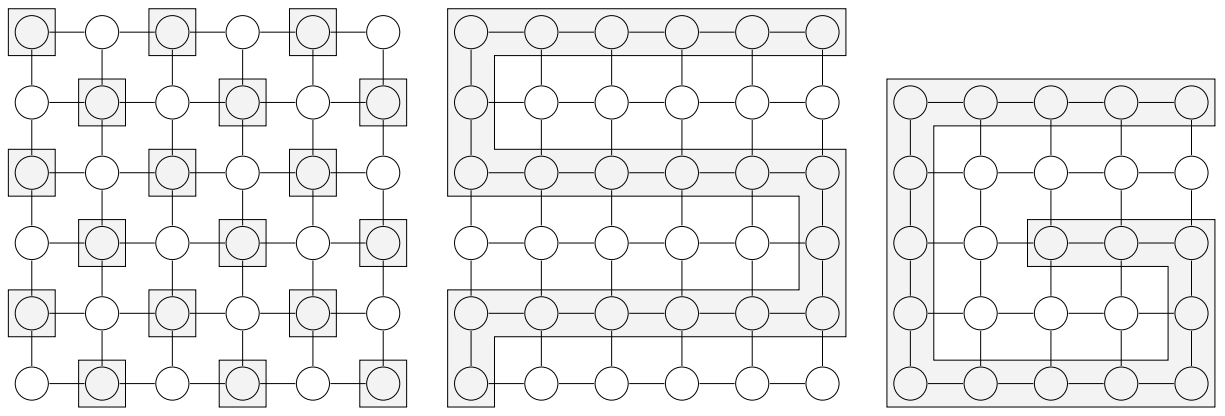

Figure 3: Alternative partitions of an MRF corresponding to data augmentation methods. Again, the shaded nodes are called $X_{\Delta_1}$. For the leftmost scheme, the elements of $\Delta_1$ are separated by $\Delta_2$, and so the conditional $p(x_{\Delta_1}|x_{\Delta_2}, y)$ is a product of the conditionals of each node. In the rightmost partition there are no unconnected subregions of either partition. The middle case is intermediate.

Smith 1990):

$$\delta_{rb}(h(X_i)) = \frac{1}{T}\sum_{t=1}^{T}\mathbb{E}[h(X_i)|\mathbf{x}_{\Delta_1}^{(t)}, \mathbf{y}]$$

where $T$ denotes the number of samples and , in this case, $i \in \Delta_2$. The alternative Monte Carlo histogram estimator is:

$$\delta_{mc}(h(X_i)) = \frac{1}{T}\sum_{t=1}^{T} h(X_i)$$

Both estimators converge to $\mathbb{E}(h(X_i)$, (Liu, Wong and Kong 1994) have proved that sampling from data augmentation is a *sufficient condition* for the Rao-Blackwellised estimator to dominate(have lower variance.) To obtain estimates of the node beliefs, we can simply choose $h$ to be the set indicator function, yielding:

$$\widehat{p}(\mathbf{x}_i|\mathbf{y}) = \frac{1}{T}\sum_{t=1}^{T} p(\mathbf{x}_i|\mathbf{x}_{\Delta_k}^{(t)}, \mathbf{y})$$

where $k = 1$ if $i \in \Delta_2$ or 2 if $i \in \Delta_1$. Rao-Blackwellised estimators are more expensive per sample as they require exact computation of the expectations. In practice, the large gain in variance reduction typically offsets this extra cost. Our experiments will show this in our setting.

As shown in Figure 3, we can partition MRFs into trees in several different ways. (We address the theoretical advantages of these and other partitioning schemes in Section 3.) Thus it is possible at each iteration of the MCMC sampler to draw a uniform random variable and choose a particular tree partition. In particular, if we consider two different tree sampling algorithms with Markov transition kernels $K_1$ and $K_2$ and stationary distribution $p(\mathbf{x}|\mathbf{y})$, then the mixture kernel $K_{mix} = \alpha K_1 + (1-\alpha)K_2$, with $0 \leq \alpha \leq 1$, also converges to $p(\mathbf{x}|\mathbf{y})$ (Tierney 1994).

# 3 THEORETICAL JUSTIFICATION

We now turn to a theoretical justification of our algorithm. It is clear that Rao-Blackwellisation reduces the variance of the estimates. However, the question of what partitions should be adopted must be addressed. As shown in Figure 3 one could partition the graph using trees or, as proposed in (Greenwood, McKeague and Wefelmeyer 1996), using a checker board. In this section, we will show that our tree partitioning scheme results in lower dependency between Markov chain samples and faster geometric convergence. To do this, we need to introduce some ideas from functional analysis.

Let $\mathbf{x}_1, \mathbf{x}_2, \ldots$ be consecutive samples of the stationary Markov chain with equilibrium distribution $\pi(\mathbf{x})$ (for notational simplicity we suppress conditioning on the data) and transition kernel $K(\mathbf{x}_k|\mathbf{x}_{k-1})$. We define the forward $F$ and backward $B$ operators on this chain as follows:

$$\begin{array}{lll} Fh(\mathbf{x}_1) & \triangleq \int h(\mathbf{y})K(\mathbf{y}|\mathbf{x}_1)d\mathbf{y} & = \mathbb{E}(h(\mathbf{x}_2)|\mathbf{x}_1) \\ Bh(\mathbf{x}_2) & \triangleq \int h(\mathbf{y})\frac{K(\mathbf{x}_2|\mathbf{y})\pi(\mathbf{y})}{\pi(\mathbf{x}_2)}d\mathbf{y} & = \mathbb{E}(h(\mathbf{x}_1)|\mathbf{x}_2) \end{array}$$

When the Markov chain is defined on a finite (discrete) space, $F$ is just a matrix. For more generality, we need to work on the Hilbert space of zero mean and finite variance functions with respect to $\pi$, denoted as:

$$L_0^2(\mathbf{x}) = \left\{ h(\mathbf{x}) : \int h^2(\mathbf{x})\pi(\mathbf{x})d\mathbf{x} < \infty, \; \int h(\mathbf{x})\pi(\mathbf{x})d\mathbf{x} = 0 \right\}$$

This is a space equipped with the inner product:

$$\langle h(\mathbf{x}), g(\mathbf{x})\rangle = \mathbb{E}_\pi(h(\mathbf{x})g(\mathbf{x}))$$

and consequently, the variance is

$$var_\pi(h(\mathbf{x})) = \langle h(\mathbf{x}), h(\mathbf{x})\rangle = \|h(\mathbf{x})\|^2$$

and the correlation coefficient is

$$corr(h(\mathbf{x}_1), g(\mathbf{x}_2)) = \left\langle \frac{h(\mathbf{x}_1)}{\|h(\mathbf{x}_1)\|}, \frac{g(\mathbf{x}_2)}{\|g(\mathbf{x}_2)\|} \right\rangle$$

The induced operator norm is defined as

$$\|F\| = \sup_{h:\|h\|=1} \|Fh\|$$

The forward and backward operators are self adjoint $\langle Fh(\mathbf{x}), g(\mathbf{x})\rangle = \langle h(\mathbf{x}), Bg(\mathbf{x})\rangle$. That is, the chain is reversible and satisfies detailed balance (Liu 2001).

Iterating the Markov chain, we have

$$F^n h(\mathbf{x}_0) = \mathbb{E}(h(\mathbf{x}_n)|\mathbf{x}_0) \quad (2)$$

$$B^n h(\mathbf{x}_n) = \mathbb{E}(h(\mathbf{x}_0)|\mathbf{x}_n) \quad (3)$$

Since $var(\mathbb{E}(h(\mathbf{x}_1)|\mathbf{x}_0)) \leq var(h(\mathbf{x}_1))$, $F$ is a contraction operator $\|Fh\| \leq \|h\|$ (its norm is bounded above by 1).

Finally, we define the maximal correlation between two random variables as follows:

$$\begin{aligned}\gamma(\mathbf{x}_2, \mathbf{x}_1) &\triangleq \sup_{g:\|g\|<\infty, h:\|h\|<\infty} corr\,(h(\mathbf{x}_2), g(\mathbf{x}_1)) \\ &= \sup_{h:\|h\|=1} \left(var\left[\mathbb{E}_\pi(h(\mathbf{x}_2)|\mathbf{x}_1)\right]\right)^{\frac{1}{2}} \\ &= \sup_{h:\|h\|=1} \|\mathbb{E}_\pi(h(\mathbf{x}_2)|\mathbf{x}_1)\| \qquad (4)\end{aligned}$$

Substituting (2) and (3) into (4) results in the following important result (Liu 2001):

**Lemma 1** $\gamma(\mathbf{x}_0, \mathbf{x}_n) = \|F^n\| = \|B^n\|$.

Note that the largest eigenvalue of $F$, when restricted to $L_0^2$, is the second largest eigenvalue of the standard Markov chain operator when restricted to $L^2$. Consequently, the reversibility of the Markov chain implies that $F$ is self-adjoint and hence its norm is equal to the second largest eigenvalue of the standard Markov chain. Since $F$ is also nonnegative, it has a decomposition $F = A^2$ where $A$ is a contracting, self-adjoint operator. We can use this to assert that the covariance between samples of the Markov chain decreases monotonically. That is,

$$cov(h(\mathbf{x}_n), h(\mathbf{x}_0)) = \langle F^n h, h\rangle = \langle A^n h, A^n h\rangle = \|A^n h\|^2$$

and by contraction, $\|A^n h\|$ decreases monotonically. This implies that algorithms with a lower operator norm will result in less correlated samples. Moreover, they will converge at a faster geometric rate (Liu et al. 1994, Lemma 12.6.3):

$$\left|\iint h(\mathbf{y}) K^n(\mathbf{y}|\mathbf{x}) P(\mathbf{x}_0) d\mathbf{y} d\mathbf{x}_0 - \mathbb{E}_\pi(h(\mathbf{x}))\right| \leq c_0 \|F\|^n \|h\|$$

Under the choice $h(\mathbf{x}) = \mathbb{I}_A(x) - \pi(A)$, where $\mathbb{I}_A(x)$ is the indicator function of the set $A$, we have:

$$|P_n(A) - \pi(A)| \leq c\|F\|^n \qquad (5)$$

That is, the Markov chain beliefs converge to the true beliefs with rate determined by $\|F\|$.

Armed with these functional analysis tools, we attack the partitioning question. In (Liu et al. 1994) this question is posed as: given random variables $U, V, W$, which of the following sampling schemes yields a better estimate?

1. $(U|V), (V|U)$
2. $(U|V, W), (V, W|U)$
3. $(U|V, W), (V|U, W), (W|U, V)$

The first two schemes are both data augmentations, but variable $W$ is "integrated out" in scheme 1. The norms of the forward operators of schemes 1 and 3 are smallest and largest respectively, so that the first scheme is favourable if the computational expense is justified. The last scheme is a general Gibbs sampler.

We also aim to choose between sampling schemes to judge which MRF partition is better, but with some important differences. In our case, it is not merely a matter of more variables being conditioned on in one case than another. The problem is that the samples are multivariate and the variables are shuffled to different phases of the sampling from one scheme to the next. An example will soon clarify clarify this.

Our comparison will focus the checker-board (CB) and two-tree sampling (TS) schemes shown in Figure 3. These two cases are extreme situations. Specifically, in the CB case, the node's neighbors render the rest of the graph conditionally independent. In Figure 3, conditioning on the "gray" set means that at a given sampling instant, each "white" node only recieves input from its "gray" neighbors and vice versa for the gray nodes. Of course as sampling proceeds, information is passed globally accross the graph but we show that this performs quite poorly. In the two-tree case, if we condition on the "gray" set, then *every* node in the "white" set is influenced by the *full gray set*.

To be precise and general, our notation must be expanded. We define the index sets of the fully-connected and checker-board schemes to be $(W_1, W_2)$, $(V_1, V_2)$ respectively (again, $W_1 \cup W_2 = V_1 \cup V_2 = I$) Our first result shows that the *variance* of the conditional expectation terms in the Rao-Blackwellised estimator is *larger* for the checker-board sampler than it is for tree sampling. Specifically,

**Proposition 1** *Given the TS and CB sampling schemes, assume without loss of generality that node i belongs to set* $W_1$ *in TS and set* $V_1$ *in CB. Then* $var(E[h(X_i)|X_{W_2}, y]) \leq var(E[h(X_i)|X_{V_2}, y])$

**Proof:** See appendix.

Proposition 1 is only applicable when a single component of a data augmentation block has information about the rest of the graph conditioned out. If we were sampling independently from the corresponding distributions, this would be a final answer to the variance reduction question, as shown in (Liu et al. 1994). However, in our case, we must consider functions of the full multivariate blocks of nodes not just of the single components $X_i$.

As a simple illustration of this additional difficulty, consider Figure 4 showing a very simple 2x2 MRF sampled using the TS and CB schemes. Adjacent to each is a corresponding "unrolled" sampling sequence showing the *spatiotemporal* relationships of the sampled variables in the usual manner of graphical models. In the ovals are the variables corresponding to the sampling blocks; the superscripts denote the *time* indices of the samples. Arrows indicate the sampling direction and reveal the "parent/child" temporal dependencies between variables. The undirected links in the ovals of the TS case reflect the spatial dependence of the variables in the blocks. In this example, our samples are multivariate. Hence, we need to compare functions of *all* the variables in a block, say $h(x_1, x_4)$ in TS against $h(x_1, x_2)$ in CB sampling. Here, there is the additional difficulty that the variables are shuffled to different times in the sampling schemes (*e.g.* $x_4$ in TS does not match $x_2$ in CB directly).

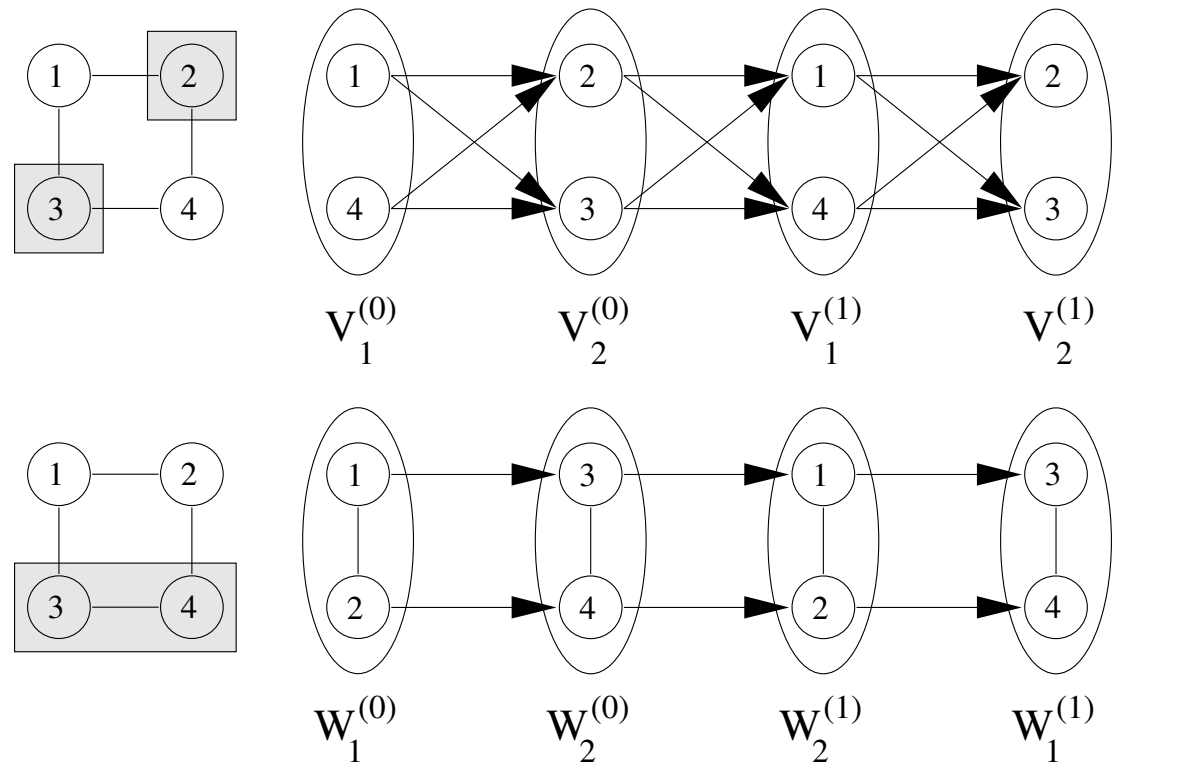

Figure 4: Illustration of chessboard (top) and fully-connected (bottom) partitioning of variables for a 2x2 MRF. In the top and bottom schemes, the shaded nodes delineate variable sets $V_2$ and $W_2$ respectively. The chains beside each show the spatiotemporal dependencies of variables in the resulting Markov chains for each scheme. Theory and experiment reveal that the fully-connected scheme yields a superior RB estimator.

Despite these additional difficulties, we show in the next theorem that the maximal correlation between functions of multivariate samples in CB is greater than the one between functions in TS.

**Theorem 1** *Let the maximal correlations between samples drawn from TS and CB be* $\gamma(X_{W_1}, X_{W_2}), \gamma(X_{V_1}, X_{V_2})$ *respectively. Then under the stationary distribution* $\gamma(X_{W_1}, X_{W_2}) \leq \gamma(X_{V_1}, X_{V_2})$

**Proof:** See appendix.

In (Liu et al. 1994) the case is made that conditioning on "more" variables (scheme 2 at the beginning of the section) can be a worse estimator, but the preceeding shows how a sampler can be better than another even though the same *number* of variables is sampled in both cases. Another interesting observation is that while the covariances of the TS sampler "instantly" involve the entire block of variables in the graph, those from CB have a sort of "propagation delay." It is easier to appreciate this by picturing a larger graph, say 10x10 or bigger. In CB, the initial conditioning set for a particular node is its 4 neighbors, and the conditioining set for *those 4 neighbors* will be the original one and the immediately surrounding region, and so on. We observe a wave-like spreading of interaction with a delay of the order of half the size of the graph before the whole block of variables is involved, and many "reflected" covariance terms generated at each iteration. This is an additional cause of higher variance with regions being conditioned out.

Finally, we state a result that may be more fundamental than that of Theorem 1. Instead of using correlations as a measure of depedency between samples (the standard approach in statistics), we propose the use of information theory measures to assess this dependency. The following theorem demonstrates how we can use information theory measures of dependency to compare MCMC sampling schemes.

**Theorem 2** *Under the stationary distribution, the* mutual information *between samples generated from CB is* larger *than that between samples from TS:*

$$I(X^{(0)}_{V_1}; X^{(0)}_{V_2}) \geq I(X^{(0)}_{W_1}; X^{(0)}_{W_2})$$

**Proof:** See appendix.

# 4 NUMERICAL RESULTS

Our experiments were aimed at demonstrating the estimator dominance result of our tree sampler (TS). We compared TS against a plain Gibbs (PG) sampler and the checker-board (CB) data augmentation mentioned in a previous section. As we predict in the theoretical discussion of Section 3, we see that for small graphs, while TS does indeed outperform CB, the gain seems somewhat incremental. For larger graphs the true merit of tree sampling becomes obvious.

We ran two sets of experiments where we applied the 3 samplers to a problem and assessed the variance or error. In the first experiment, for a small (10x10) MRF with isotropic potentials and each node having between 10 and 15 possible states, we approximated the expected value of each node using PG, CB and TS with the goal of assessing *variance* of the estimators. Each sampler was given 500 trials of 1200 iterations. The empirical variance of the mean estimator of each node was calculated over the trials, and records of the estimate against computation were obtained. Figure 5 shows a node-by-node comparison of the variances of the mean estimators of the samplers *adjusted for computational time.* To obtain this measure for "value for money" we penalized slow estimators by multiplying the raw estimator variance by the time factor by which they were greater than the fastest scheme. Thus, the PG was not penalized (fastest), and TS was penalized more than CB. Nonetheless our method is an improvement over the CB despite the penalty factor. Note how the variance of ours seems to "envelop" the chessboard sampler, precisely as predicted by the bounds. Per unit computation time, CB achieved a variance reduction factor of 10.41 over PG, but our sampler reduced it by 17.18. In Figure 6, we plot the estimated mean against computation time for a particular node in the MRF; the median estimate over the trials is plotted. A vital aspect to note are the standard deviation error bars accross trials. It is clear that the PG and, to a certain extent CB, are subject to wild swings in estimation. The thick error bars are from TS, showing how much more *stable* an estimator TS produces, yielding estimates that are more trustworthy as evinced by the narrow range across trials. We also notice in passing that we found that loopy belief propagation failed to converge for several random instance of these small MRFS.

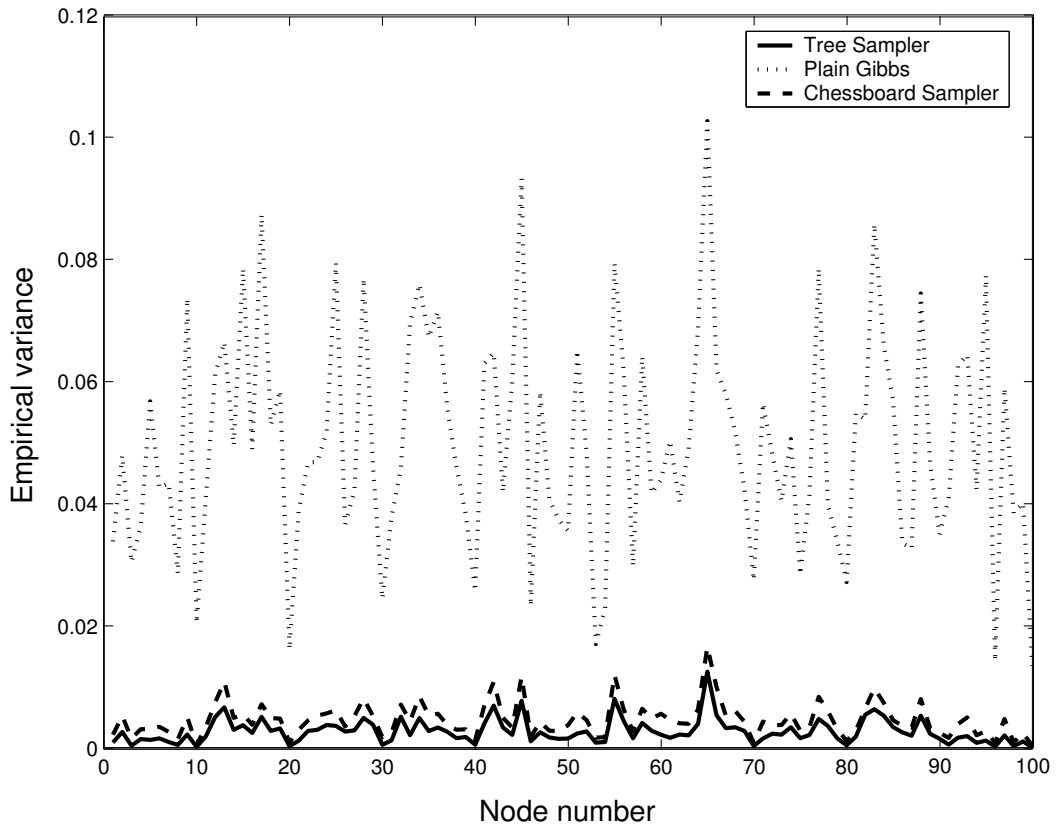

Figure 5: Computation time-adjusted variance estimates of a 10x10 MRF as a function of the node for the 3 samplers discussed in the text. Note the upper-bounding effect; the "shapes" of the graphs are similar but the magnitudes are scaled, which is an expected consequence of Rao-Blackwellisation.

Our next experiment set was the classic "reconstruction" of states from noisy observations. We used a $50 \times 50$ pixel "patch" image (consisting of shaded, rectangular regions) with an isotropic 11-state prior model. Noise was added by randomly flipping states. Each sampler was run for 1000 iterations on each of 50 separate trials. An important aspect to assess is the *sensitivity* of the estimator, that is, is our good estimate a matter of luck or is it robust over trials? The plot in Figure 7 shows the median reconstruction error as a function of *computation time* showing that the gain is considerable. In fact in this case, the CB sampler is hardly distinguishable from PG, again a predictable consequence of the theoretical considerations. For larger graphs, far from expecting any kind of breakdown of these results, we *predict that the difference will become even sharper.* The error bars show that the low reconstruction error of our sampler is highly robust across trials compared to that of PG and CB. We also ran loopy on these images, which took about the same number of iterations (around 30 passes through the MRF) to achieve the same error reduction as our method, suggesting that our method might be computationally comparable to loopy but guaranteed to converge. It is very important to realize that the gain is *not* merely due to "blocking"; the CB sampler *is also* a 2-block Rao-Blackwellised scheme, but *does not take advantage of RB as well.*

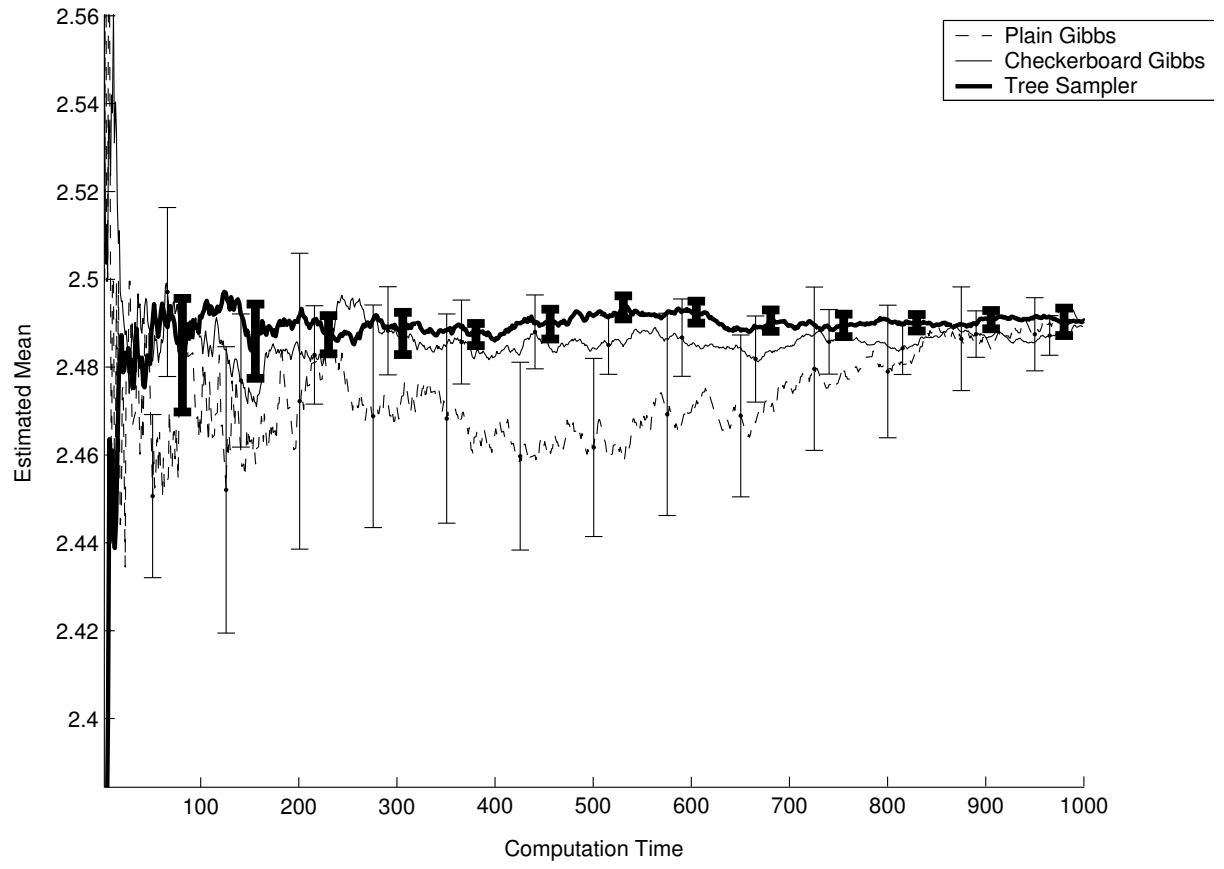

Figure 6: Convergence to the estimated mean of PG, CB, and our sampler for a specific node of a $10 \times 10$ MRF, for which LBP failed. The error bars centered on each line represent the standard deviation of the estimates across trials. PG has the widest bars and TS, shown with thick lines, has the narrowest. The plots are the *medians* of the trials. Note the considerable increase in stability and variance reduction in the case of our sampler, which outperforms the improvements afforded by CB. The error ranges for our sampler and CB always overlap, while PG sometimes fluctuates out of range.

# 5 CONCLUSION

We proposed a new tree sampling algorithm for MRFs. We addressed the issue of comparing different partitioning schemes, ranging from checker-board (CB) sampling to two-tree sampling (TS). These two algorithms are extremes in a scale where TS exploits larger analytical coverage of the graph. We proved two fundamental theorems. Theorem 1 showed that the maximal correlation between samples drawn from the CB sampler is larger than the one for TS. This implies that TS exhibits a faster geometric convergence rate and less correlation between samples. Theorem 2 showed that information theory measures can be applied to the analysis of MCMC algorithms. In particular, it showed that he *mutual information* between samples from CB is higher than the one for TS. Our experimental results confirmed these assertions.

# APPENDIX: PROOFS

**Proof of Proposition 1:** *Recall that for any random variables* $A, B, C$, $E[A|B] = E[E[A|B,C]|B]$ *so that* $var(E[A|B]) \leq var(E[A|B,C])$. *Our proof follows from this and the* spatial *Markov properties of the graph. We note that for the TS case, the expectation can be written as*

$$E[h(X_i)|x_{W_2}, y] = E[E[h(X_i)|X_{V_2}, X_{W_2}, y]|X_{W_2}, y]$$

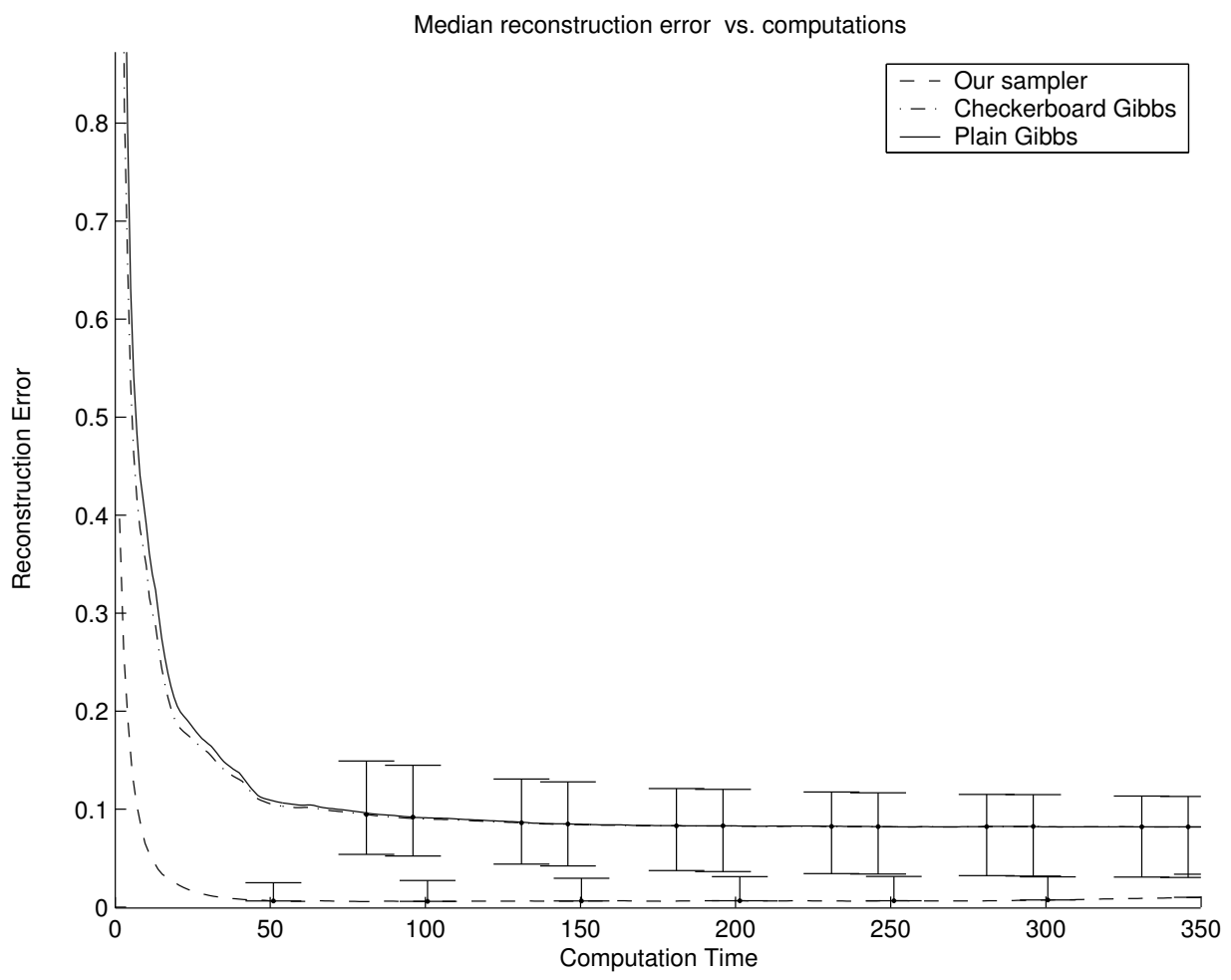

Figure 7: Reconstruction error against computation time for a $50 \times 50$ pixel 11-state image corrupted with noise. The 3 plots show the median reconstruction error history of PG, CB and our sampler over 50 runs. The bars represent the standard deviation of the error across these runs. Clearly, aside from achieving a lower error, our sampler *is more robust*, that is, consistently achieves these results. The gain of CB over PG in this case is negligible, again predictable from the theory of Rao-Blackwellisation

*But by conditional independence,*

$$E[h(X_i)|x_{V_2}, X_{W_2}, y] = E[h(X_i)|X_{V_2}, y]$$

*which is precisely the form of the expectation terms of the CB sampler. Thus the TS expectations correspond to an additional conditioning step over the CB expectations, and by the above relation for conditioning, the variance result follows.*

**Proof of Theorem 1:** *We show this for the small 2x2 MRF shown in Figure 4; the extension to larger MRFs is inductive. The proof depends on the decomposition of the transition kernel in both cases; it will turn out that we can do a term-by-tem analysis of the resulting decompositions and show that some of the TS terms are CB terms with additional integration.*

*Let the respective joint/conditional probabilities under CB and TS be $p_{CB}$ and $p_{TS}$. For the CB sampler, the one-step transition kernel of the* joint *chain is:*

$$\begin{aligned} K_{CB}(x^{(0)}, x^{(1)}) &= p_{CB}(x_1^{(1)}|x_2^{(0)}, x_3^{(0)}) p_{CB}(x_2^{(1)}|x_1^{(1)}, x_4^{(1)}) \\ &\times \; p_{CB}(x_3^{(1)}|x_1^{(1)}, x_4^{(1)}) p_{CB}(x_4^{(1)}|x_2^{(0)}, x_3^{(0)}) \end{aligned}$$

*while for the TS sampler, the kernel is:*

$$\begin{aligned} K_{TS}(x^{(0)}, x^{(1)}) &= p_{TS}(x_1^{(1)}|x_3^{(0)}, x_4^{(0)}) p_{TS}(x_2^{(1)}|x_1^{(1)}, x_4^{(0)}) \\ &\times \; p_{TS}(x_3^{(1)}|x_1^{(1)}, x_2^{(1)}) p_{TS}(x_4^{(1)}|x_2^{(1)}, x_3^{(1)}) \end{aligned}$$

*Using the reversibility of the augmentation chains, the assumption of stationarity, and the spatial properties of the graph, we see that* all *of the above conditional distributions are conditionals from $\pi$, the stationary distribution. For example $p_{CB}(x_2^{(1)}|x_1^{(1)}, x_4^{(1)}) = \pi(x_2|x_1, x_4)$. Also note that $p_{CB}(x_2^{(1)}|x_1^{(1)}, x_4^{(1)}) = p_{TS}(x_2^{(1)}|x_1^{(1)}, x_4^{(0)}) = \pi(x_2|x_1, x_4)$ and $p_{CB}(x_4^{(1)}|x_2^{(0)}, x_3^{(0)}) = p_{TS}(x_4^{(1)}|x_2^{(1)}, x_3^{(1)}) = \pi(x_4|x_2, x_3)$. By applying the spatial Markov property in "reverse,"*

$$\pi(x_1^{(1)}|x_2^{(0)}, x_3^{(0)}) = \pi(x_1^{(1)}|x_2^{(0)}, x_3^{(0)}, x_4^{(0)})$$

*and*

$$\pi(x_3^{(1)}|x_1^{(1)}, x_4^{(1)}) = \pi(x_3^{(1)}|x_1^{(1)}, x_2^{(1)}, x_4^{(1)})$$

*If we define the following functions (for conciseness domain variables are omitted.)*

$$\begin{aligned} j_1 &= \pi(x_1^{(1)}|x_2^{(0)}, x_3^{(0)}, x_4^{(0)}) \\ j_2 &= \pi(x_2^{(1)}|x_1^{(1)}, x_4^{(1)}) \\ j_3 &= \pi(x_3^{(1)}|x_1^{(1)}, x_2^{(1)}, x_4^{(1)}) \\ j_4 &= \pi(x_4^{(1)}|x_2^{(0)}, x_3^{(0)}) \end{aligned}$$

*and:*

$$\begin{aligned} k_1 &= \int j_1 \pi(x_2^{(0)}|x_3^{(0)}, x_4^{(0)}) dx_2^{(0)} \\ k_2 &= \pi(x_2^{(1)}|x_1^{(1)}, x_4^{(0)}) \\ k_3 &= \int j_3 \pi(x_4^{(1)}|x_1^{(1)}, x_2^{(1)}) dx_4^{(0)} \\ k_4 &= \pi(x_4^{(1)}|x_2^{(1)}, x_3^{(1)}) \end{aligned}$$

*then we can write the transition kernel of CB as:*

$$K_{CB}(x^{(0)}, x^{(1)}) = j_1 j_2 j_3 j_4 \quad (6)$$

*and that of TS as:*

$$K_{TS}(x^{(0)}, x^{(1)}) = k_1 k_2 k_3 k_4 \quad (7)$$

*If we compare the conditional expectation operators acting on a given function $g(x)$ $(= g(x_1, x_2, x_3, x_4))$ under the two schemes, we have:*

$$E_{CB}[g(X^{(1)})|x^{(0)}] = \int j_1 \int j_2 \int j_3 \int j_4 g(x^{(1)}) dx_{1:4}^{(0)}$$

*and:*

$$E_{TS}[g(X^{(1)})|x^{(0)}] = \int k_1 \int k_2 \int k_3 \int k_4 g(x^{(1)}) dx_{1:4}^{(0)}$$

*Note that the expression for the TS is similar to CB's but the conditional distributions of $X_1$ and $X_3$ ($k_1$ and $k_3$) are the* integrated *expressions given in (6). Clearly, as integration proceeds "outward" from $X_4$ to $X_1$ in both cases, the variance of the argument function is reduced by the same amount for the $X_2$ and $X_4$ integrations (it is the same operator in both cases due to the distributions being the same) but is reduced more in TS for the $X_1$ and $X_3$ operators due to the extra integration. It can be verified that this effect becomes progressively more pronounced as the graph gets* *larger, with a greater proportion of the terms having extra integrations, consistent with our experimental results. Thus the norm of the CB expectation operator is larger than TS's, and by Theorem 3.2 of (Liu et al. 1994), the result on maximal correlation follows.*

**Proof of Theorem 2:** *Let the conditional entropies be*

$$\begin{aligned} H_{TS} &\triangleq H(X_{W_1}^{(1)}, X_{W_2}^{(1)} | X_{W_1}^{(0)}, X_{W_2}^{(0)}) \\ H_{CB} &\triangleq H(X_{V_1}^{(1)}, X_{V_2}^{(1)} | X_{V_1}^{(0)}, X_{V_2}^{(0)}) \end{aligned}$$

*Using the same decomposition of the probabilities and stationarity/reversibility arguments, the spatial Markov properties used to prove Theorem 1, and the fact that conditioning reduces entropy, it can be shown that*

$$H_{TS} \geq H_{CB}$$

*Using the* temporal *Markov properties,*

$$\begin{aligned} H_{CB} = & H(X_1^{(1)} | X_2^{(0)}, X_3^{(0)}) + H(X_2^{(0)} | X_1^{(0)}, X_4^{(0)}) + \dots \\ & H(X_3^{(0)} | X_1^{(0)}, X_4^{(0)}) + H(X_4^{(1)} | X_2^{(0)}, X_3^{(0)}) \end{aligned}$$

*with an analogous expansion for $H_{TS}$. Now since the* joint *entropy of* both *schemes is the same, and*

$$\begin{aligned} I(X_{W_1}^{(0)}; (X_{W_1}^{(1)}, X_{W_2}^{(0)})) &= I(X_{W_1}^{(0)}; X_{W_2}^{(0)}) \\ I(X_{V_1}^{(0)}; (X_{V_1}^{(1)}, X_{V_2}^{(0)})) &= I(X_{V_1}^{(0)}; X_{V_2}^{(0)}) \end{aligned}$$

*by the definition of joint entropy we obtain*

$$H_{TS} + I(X_{W_1}^{(0)}; X_{W_2}^{(0)}) = H_{CB} + I(X_{V_1}^{(0)}; X_{V_2}^{(0)})$$

*By the stationarity of the chain, the mutual information result holds for* any *two successive samples drawn from TS and CB.*

## References

Besag, J. E. (1974). Spatial interaction and the statistical analysis of lattice systems, *Journal of the Royal Statistical Society B* **36**: 192–236.

Besag, J. E. (1986). On the statistical analysis of dirty pictures, *Journal of the Royal Statistical Society B* **48**(3): 259–302.

Bidyuk, B. and Dechter, R. (2003). An empirical study of w-cutset sampling for Bayesian networks, *UAI*.

Carter, C. K. and Kohn, R. (1994). On Gibbs sampling for state space models, *Biometrika* **81**(3): 541–553.

Casella, G. and Robert, C. P. (1996). Rao-Blackwellisation of sampling schemes, *Biometrika* **83**(1): 81–94.

Doucet, A., de Freitas, N., Murphy, K. and Russell, S. (2000). Rao-Blackwellised particle filtering for dynamic Bayesian networks, *in* C. Boutilier and M. Godszmidt (eds), *Uncertainty in Artificial Intelligence*, Morgan Kaufmann Publishers, pp. 176–183.

Gelfand, A. E. and Smith, A. F. M. (1990). Sampling-based approaches to calculating marginal densities, *Journal of the American Statistical Association* **85**(410): 398–409.

Geman, S. and Geman, D. (1984). Stochastic relaxation, Gibbs distributions and the Bayesian restoration of images, *IEEE Transactions on Pattern Analysis and Machine Intelligence* **6**(6): 721–741.

Greenwood, P., McKeague, I. and Wefelmeyer, W. (1996). Outperforming the Gibbs sampler empirical estimator for nearest neighbor random fields, *Annals of Statistics* **24**: 1433–1456.

Kumar, S. and Hebert, M. (2003). Discriminative fields for modeling spatial dependencies in natural images, *in proc. advances in Neural Information Processing Systems (NIPS)*.

Li, S. Z. (2001). *Markov random field modeling in image analysis*, Springer-Verlag.

Liu, J. S. (ed.) (2001). *Monte Carlo Strategies in Scientific Computing*, Springer-Verlag.

Liu, J., Wong, W. H. and Kong, A. (1994). Covariance structure of the Gibbs sampler with applications to the comparisons of estimators and augmentation schemes, *Biometrika* **81**(1): 27–40.

McCallum, A., Rohanimanesh, K. and Sutton, C. (2003). Dynamic conditional random fields for jointly labeling multiple sequences, *NIPS Workshop on Syntax, Semantics and Statistics*.

Paskin, M. (2003). Sample propagation, *Advances in Neural Information Processing System*.

Pearl, J. (1987). Evidential reasoning using stochastic simulation, *Artificial Intelligence* **32**: 245–257.

Robert, C. P. and Casella, G. (1999). *Monte Carlo Statistical Methods*, Springer-Verlag, New York.

Swendsen, R. H. and Wang, J. S. (1987). Nonuniversal critical dynamics in Monte Carlo simulations, *Physical Review Letters* **58**(2): 86–88.

Tierney, L. (1994). Markov chains for exploring posterior distributions, *The Annals of Statistics* **22**(4): 1701–1762.

Wainwright, M., Jaakkola, T. and Willsky, A. (2003). Tree-reweighted belief propagation and approximate ML estimation by pseudo-moment matching, *AI-STATS*, Florida, USA.

Wilkinson, D. J. and Yeung, S. K. H. (2001). Conditional simulation from highly structured gaussian systems, with application to blocking-MCMC for the Bayesian analysis of very large linear models, *Statistics and Computing* **11**: To appear.

Yedidia, J. S., Freeman, W. T. and Weiss, Y. (2000). Generalized belief propagation, *in* S. Solla, T. Leen and K.-R. Müller (eds), *Advances in Neural Information Processing Systems 12*, MIT Press, pp. 689–695.